# Geometric Correlation between Dirac Equation and Yang-mills Equation/ Maxwell Equation

Xue-Gang Yu[*]

*Department of Physics, College of Science, Tianjin University of Commerce,*

*Tianjin 300134, People's Republic of China*

**Abstract:** The problem about geometric correspondence of Dirac particle and contain quality item of Yang-Mills equation has always not been solved.This paper introduced the hyperbolic imaginary unit in Minkowski space, established a classes of Dirac wave equations with t'Hooft matrices.In lightlike region of Minkowski space,we can discuss the hermitian conjugate transformation of Dirac positive particle and antiparticle, find the space-time corresponding points of Dirac particle,and draw Feynman clip-art though the geometrical relation between timelike region and lightlike region.The coupling of motion equation of Dirac positive particle and antiparticle can get Klein-Gordon equation, when it reach classical approximate we can get Schrodinger equation,and this illustrated that $\pi$ meson or $\mu$ meson may be composite particle. Using the relation of timelike region and lightlike region in Minkowski momentum space to renormalize the rest mass of particles,we can describe the geometric relation between rest mass and electromagnetic mass of particles. Then, we can elicit the Yang-Mills equation with electromagnetic mass through four Dirac wave equations with the hermitian conjugate transformation relation, and further launch the common forms of Maxwell equations.

**Subjects:** Mathematical physics(math-ph); High energy physics-Theory(hep-th)

## 1. Introduction

Introduce the imaginary unit $j(j^2 = 1, j^* = -j)$ in the Clifford geometric algebra, the corresponding space is Minkowski space (*1*). We define linear quaternions as follows:

$$\begin{cases} X_\mu = ct + j\bar{r} \\ X_\mu^* = ct - j\bar{r} \end{cases} \quad (1.1)$$

where $t$ is time, $\bar{r}$ is three-dimensional space vector. Take the inner product of Eq. (1.1) to obtain the space-time interval invariant

$$X_\mu^* X_\mu = c^2 t^2 - r^2 = R^2 \quad (1.2)$$

$R^2 > 0$ is timelike region in Minkowski space, $R^2 = 0$ is lightlike region, and

---

[*] E-mail: yxgsl@tjcu.edu.cn

$R^2 < 0$ is spacelike region. Let four-dimensional velocity (2)

$$\varpi_\mu = \frac{dX_\mu}{d\tau} = \frac{1}{\alpha}(c + j\vec{v}) \qquad (1.3)$$

where, $\tau = \alpha t$ is self-time, $\alpha = (1 - \frac{v^2}{c^2})^{\frac{1}{2}}$ is relativity factor. Take the inner product of Eq. (1.3), therefore

$$\varpi_\mu^* \varpi_\mu = c^2 \qquad (1.4)$$

Let four-dimensional momentum

$$P_\mu = m_0 \varpi_\mu = \frac{E}{c} + j\vec{p} \qquad (1.5)$$

where $m_0$ is rest mass. Take the inner product of Eq. (1.5), energy-momentum relation is obtained as follows (3):

$$P_\mu^* P_\mu = \frac{E^2}{c^2} - p^2 = m_0^2 c^2 \qquad (1.6)$$

The momentum and energy operators can be written as, respectively

$$\hat{\vec{p}} = j\hbar\vec{\nabla}, \hat{E} = -j\hbar\frac{\partial}{\partial t} \qquad (1.7)$$

The four-dimensional momentum operators are:

$$\begin{cases} \hat{P}_\mu = \hat{E} + j\hat{\vec{p}} = -j\hbar\square^+ = -j\hbar(\frac{\partial}{\partial X_\mu})^+ & (1.8a) \\ \hat{P}_\mu^+ = \hat{E}^+ - j\hat{\vec{p}}^+ = j\hbar\square = j\hbar(\frac{\partial}{\partial X_\mu}) & (1.8b) \end{cases}$$

where

$$\begin{cases} \frac{\partial}{\partial X_\mu} = \frac{\partial}{\partial ct} + j\vec{\nabla} \\ (\frac{\partial}{\partial X_\mu})^+ = \frac{\partial}{\partial ct} - j\vec{\nabla} \end{cases} \qquad (1.9)$$

Take the inner product algorithm of linear quaternions in the hyperbolic complex domain

$$<X|Y> = \sum_{i,k=1}^{4} x_i^* y_k \qquad (1.10)$$

When $X = Y$, as a particular application, the contraction of Eq. (1.10) may be written

as follows

$$<X|X> = \sum_{i=1}^{4} x_i^* x_i \quad (1.11)$$

According to the algorithm of Eq. (1.11), take the inner product of Eq.(1.8) to deduce the operator expression

$$\hbar^2 \nabla^2 - \frac{\hbar^2}{c^2} \frac{\partial^2}{\partial t^2} - m_0^2 c^2 = 0 \quad (1.12)$$

## 2. The four-dimensional hyperbolic Dirac wave equations

Introduce the second-order operators in hyperbolic complex space (different from Pauli operators)

$$\sigma_x = \begin{pmatrix} & 1 \\ 1 & \end{pmatrix}, \quad \sigma_y = \begin{pmatrix} 1 & \\ & -1 \end{pmatrix}, \quad \sigma_z = \begin{pmatrix} & -1 \\ 1 & \end{pmatrix} \quad (2.1)$$

They satisfy the relations

$$\sigma_i^+ \sigma_i = \sigma_i \sigma_i^+ = 1 \ (i = 1,2,3) \quad (2.2)$$

or

$$[\sigma_i^+ \sigma_k]_+ = \sigma_i^+ \sigma_k + \sigma_k \sigma_i^+ = 2\delta_{ik} \quad (2.3)$$

where $i, k = 1, 2, 3$. In hyperbolic complex space, take the fourth-order matrices $\gamma_\mu$

$$\gamma_1 = \begin{pmatrix} & -\sigma_x \\ \sigma_x & \end{pmatrix} = \begin{bmatrix} & & & -1 \\ & & -1 & \\ & 1 & & \\ 1 & & & \end{bmatrix} \quad (2.4a)$$

$$\gamma_2 = \begin{pmatrix} & -\sigma_y \\ \sigma_y & \end{pmatrix} = \begin{bmatrix} & & & -1 \\ & & 1 & \\ 1 & & & \\ & -1 & & \end{bmatrix} \quad (2.4b)$$

$$\gamma_3 = \begin{pmatrix} \sigma_z & \\ & \sigma_z \end{pmatrix} = \begin{bmatrix} & -1 & & \\ 1 & & & \\ & & & -1 \\ & & 1 & \end{bmatrix} \quad (2.4c)$$

$$\gamma_4 = \gamma_0 = \begin{bmatrix} 1 & & & \\ & 1 & & \\ & & 1 & \\ & & & 1 \end{bmatrix} \qquad (2.4d)$$

where, $\gamma_1, \gamma_2, \gamma_3$ are the t'Hooft matrices and $\gamma_4 = \gamma_0 = I$ is a unit matrix. We can verify that Eqs. (2.4) fulfill

$$\gamma_i^+ = -\gamma_i, \quad \gamma_4^+ = \gamma_4 \qquad (2.5)$$

$$\left(\gamma_\mu \frac{\partial}{\partial X_\mu}\right)^+ = \gamma_\mu \frac{\partial}{\partial X_\mu} \qquad (2.6)$$

$$\gamma_\mu^+ \gamma_\mu = \gamma_\mu \gamma_\mu^+ = 1 \qquad (2.7)$$

$$\gamma_i \gamma_i = -1 (i = 1,2,3) \qquad (2.8)$$

$$\gamma_\mu^+ \gamma_\nu + \gamma_\nu^+ \gamma_\mu = 2\delta_{\mu\nu} \qquad (2.9)$$

Here the fourth-order matrices are greatly different from the traditional ones. They satisfy the following multiplication table:

Table 1. The multiplication table of $\gamma_\mu (\gamma_1, \gamma_2, \gamma_3, \gamma_0)$

|            | $\gamma_0$  | $\gamma_1$   | $\gamma_2$   | $\gamma_3$   |
|------------|-------------|--------------|--------------|--------------|
| $\gamma_1$ | $\gamma_1$  | $-\gamma_0$  | $-\gamma_3$  | $\gamma_2$   |
| $\gamma_2$ | $\gamma_2$  | $\gamma_3$   | $-\gamma_0$  | $-\gamma_1$  |
| $\gamma_3$ | $\gamma_3$  | $-\gamma_2$  | $\gamma_1$   | $-\gamma_0$  |

Table 1 shows that

$$\gamma_1 \gamma_2 = -\gamma_3, \gamma_2 \gamma_1 = \gamma_3, \gamma_2 \gamma_3 = -\gamma_1, \gamma_3 \gamma_2 = \gamma_1, \gamma_3 \gamma_1 = -\gamma_2, \gamma_1 \gamma_3 = \gamma_2 \qquad (2.10)$$

or

$$\gamma_i \gamma_j = -\varepsilon_{ijk} \gamma_k (i, j, k = 1,2,3) \qquad (2.11)$$

$$(\gamma_i \gamma_j + \gamma_j \gamma_i) = 0 (i \neq j) \qquad (2.12)$$

and

$$\gamma_1 \gamma_2 \gamma_3 = \gamma_0 \qquad (2.13)$$

Therefore, $\gamma_1, \gamma_2, \gamma_3, \gamma_0$ are quaternions, while traditional $\gamma$ function belongs to sedenion. The former has higher symmetry than the latter.

From Eqs. (1.5), (1.6) and (1.8), take four-dimensional hyperbolic Dirac equation

$$\gamma_\mu j\hbar \frac{\partial}{\partial X_\mu} \phi = m_0 \phi \qquad (2.14)$$

Then taking Hermitian conjugate of Eq. (2.14), we obtain

$$-\phi^+ j\hbar (\frac{\partial}{\partial X_\mu})^+ \gamma_\mu^+ = m_0 \phi^+ \qquad (2.15)$$

Eq. (2.14) can be expanded as

$$\begin{cases} -\dfrac{\partial}{\partial x}\phi_4 - \dfrac{\partial}{\partial y}\phi_3 - \dfrac{\partial}{\partial z}\phi_2 + \left( j\dfrac{\partial}{\partial ct} - \dfrac{m_0 c}{\hbar} \right)\phi_1 = 0 & (2.16a) \\[2mm] -\dfrac{\partial}{\partial x}\phi_3 + \dfrac{\partial}{\partial y}\phi_4 + \dfrac{\partial}{\partial z}\phi_1 + \left( j\dfrac{\partial}{\partial ct} - \dfrac{m_0 c}{\hbar} \right)\phi_2 = 0 & (2.16b) \\[2mm] \dfrac{\partial}{\partial x}\phi_2 + \dfrac{\partial}{\partial y}\phi_1 - \dfrac{\partial}{\partial z}\phi_4 + \left( j\dfrac{\partial}{\partial ct} - \dfrac{m_0 c}{\hbar} \right)\phi_3 = 0 & (2.16c) \\[2mm] \dfrac{\partial}{\partial x}\phi_1 - \dfrac{\partial}{\partial y}\phi_2 + \dfrac{\partial}{\partial z}\phi_3 + \left( j\dfrac{\partial}{\partial ct} - \dfrac{m_0 c}{\hbar} \right)\phi_4 = 0 & (2.16d) \end{cases}$$

Eq. (2.15) may be also separated as

$$\begin{cases} -\dfrac{\partial}{\partial x}\phi_4^+ - \dfrac{\partial}{\partial y}\phi_3^+ - \dfrac{\partial}{\partial z}\phi_2^+ - \left( j\dfrac{\partial}{\partial ct} + \dfrac{m_0 c}{\hbar} \right)\phi_1^+ = 0 & (2.17a) \\[2mm] -\dfrac{\partial}{\partial x}\phi_3^+ + \dfrac{\partial}{\partial y}\phi_4^+ + \dfrac{\partial}{\partial z}\phi_1^+ - \left( j\dfrac{\partial}{\partial ct} + \dfrac{m_0 c}{\hbar} \right)\phi_2^+ = 0 & (2.17b) \\[2mm] \dfrac{\partial}{\partial x}\phi_2^+ + \dfrac{\partial}{\partial y}\phi_1^+ - \dfrac{\partial}{\partial z}\phi_4^+ - \left( j\dfrac{\partial}{\partial ct} + \dfrac{m_0 c}{\hbar} \right)\phi_3^+ = 0 & (2.17c) \\[2mm] \dfrac{\partial}{\partial x}\phi_1^+ - \dfrac{\partial}{\partial y}\phi_2^+ + \dfrac{\partial}{\partial z}\phi_3^+ - \left( j\dfrac{\partial}{\partial ct} + \dfrac{m_0 c}{\hbar} \right)\phi_4^+ = 0 & (2.17d) \end{cases}$$

Eqs. (2.16) and (2.17) are the spin-dependent hyperbolic Dirac equations, which are double times of the number of traditional Dirac equations, and form octet Dirac particles. It is different from the traditional Dirac equations that each hyperbolic Dirac equation consists of four components of the wave function, while a traditional one only contains three components. The hyperbolic Dirac equation is similar to Majovana representation. If Eq. (2.16) denotes the particle equation, then Eq. (2.17) shows of the antiparticle equation, which are mutual Hermitian conjugate. Moreover, the wave functions of the particles and antiparticles are independent and of Hermitian conjugate each other.

## 3. Covariance of the Hyperbolic Dirac Equations

Lorentz group is a special unitary group $SU(n)$ in hyperbolic complex spaces. Lorentz transformation may be written as (4)

$$X_\mu^{'} = \alpha_{\mu\nu} X_\nu \tag{3.1}$$

where

$$\alpha_{\mu\gamma}^+ \alpha_{\mu\gamma} = \alpha_{\mu\gamma} \alpha_{\mu\gamma}^+ = I \tag{3.2}$$

Under the Lorentz transformation, let the transformation of the wave function be

$$\phi^{'}(X^{'}) = \Lambda \phi(X) \tag{3.3}$$

where $X$ is a four-dimensional function, and $\Lambda$ satisfies

$$\Lambda^{-1}\Lambda = \Lambda\Lambda^{-1} = 1 \tag{3.4}$$

Take the four-dimensional gradient operator with property

$$\frac{\partial}{\partial X_\mu^{'}} = \alpha_{\mu\nu}^+ \frac{\partial}{\partial X_\nu}, \quad \frac{\partial}{\partial X_\nu} = \alpha_{\mu\nu} \frac{\partial}{\partial X_\mu^{'}} \tag{3.5}$$

Substituting Eqs.(3.3) and (3.4) into Eq.(2.14),

$$\left( j\hbar \gamma_\mu \alpha_{\mu\nu} \frac{\partial}{\partial X_\nu^{'}} - m_0 \right) \Lambda^{-1} \phi^{'}(x^{'}) = 0 \tag{3.6}$$

And multiplying Eq. (3.6) by $\Lambda$ from the left, we obtain

$$\left( j\hbar \Lambda \gamma_\mu \alpha_{\mu\nu} \Lambda^{-1} \frac{\partial}{\partial X_\nu^{'}} - m_0 \right) \phi^{'}(x^{'}) = 0 \tag{3.7}$$

Let $\Lambda \gamma_\mu \alpha_{\mu\nu} \Lambda^{-1} = \gamma_\nu$, then from Eq. (3.4) we have

$$\gamma_\mu \alpha_{\mu\nu} = \Lambda^{-1} \gamma_\nu \Lambda \tag{3.8}$$

Thus Eq. (3.7) may be turned into

$$\left( j\hbar \gamma_\nu \frac{\partial}{\partial X_\nu^{'}} - m_0 \right) \phi^{'}(x^{'}) = 0 \tag{3.9}$$

Eqs. (3.9) and (2.14) have the same form. If $\Lambda$ can be found, then it can be verified that Dirac equations have Lorentz invariance in hyperbolic complex spaces.

Make the infinitesimal Lorentz transformation to Eq. (2.14) in four-dimensional space-time. Choose

$$\alpha_{\mu\nu} = I + j\varepsilon_{\mu\nu} \tag{3.10}$$

where $I$ is a unit matrix. The invariant of space-time interval fulfills

$$\alpha_{\mu\nu}^+ \alpha_{\mu\nu} = (I - j\varepsilon_{\mu\nu}^+)(I + j\varepsilon_{\mu\nu}) = I + j(\varepsilon_{\mu\nu} - \varepsilon_{\mu\nu}^+) + O(\varepsilon_{\mu\nu}^2) = I \tag{3.11}$$

i.e., $|\varepsilon_{\mu\nu} - \varepsilon_{\mu\nu}^+|$ is small. Let

$$\begin{cases} \Lambda = I + \dfrac{1}{2} j\varepsilon_{\mu\nu} & (3.12a) \\ \Lambda^{-1} = \Lambda^+ = I - \dfrac{1}{2} j\varepsilon_{\mu\nu}^+ & (3.12b) \end{cases}$$

which satisfy the condition of Eq. (3.4), also

$$\Lambda^{-1}\gamma_\mu \Lambda = \left(I - \frac{1}{2} j\varepsilon_{\mu\nu}^+\right)\gamma_\mu \left(I + \frac{1}{2} j\varepsilon_{\mu\nu}\right) = \gamma_\mu - \frac{1}{2} j\varepsilon_{\mu\nu}^+ \gamma_\mu + \frac{1}{2} j\gamma_\mu \varepsilon_{\mu\nu} + O(\varepsilon_{\mu\nu}^2)$$

If we can find $-\varepsilon_{\mu\nu}^+ \gamma_\mu = \gamma_\mu \varepsilon_{\mu\nu}$ or

$$\gamma_\mu \varepsilon_{\mu\nu} + \varepsilon_{\mu\nu}^+ \gamma_\mu = 0 \tag{3.13}$$

then

$$\Lambda^{-1}\gamma_\mu \Lambda = \gamma_\mu + j\gamma_\mu \varepsilon_{\mu\nu} = \gamma_\mu (I + j\varepsilon_{\mu\nu}) = \gamma_\mu \alpha_{\mu\nu} \tag{3.14}$$

It can be proved that the Dirac equations are of Lorentz invariance in hyperbolic complex spaces. Such that

$$\varepsilon_{\mu\nu} = \begin{bmatrix} & & & 1 \\ & & -1 & \\ & 1 & & \\ -1 & & & \end{bmatrix} \tag{3.15}$$

Substitute Eqs. (3.15) and (2.4) into Eq. (3.13), then derive that Eq. (3.9) is of Lorentz invariance.

## 4. The Hermitian Conjugate Transformation of Particles and Antiparticles

Equation (2.15) may be written as

$$\phi^+ \left[ j\hbar \left( \frac{\partial}{\partial X_\mu} \right)^+ \gamma_\mu^+ + m_0 \right] = 0 \tag{4.1}$$

We multiply Eq. (2.14) from the left by $\phi^+$, Eq. (4.1) from the right by $\phi$, and add. Then

$$\phi^+ \gamma_\mu \frac{\partial}{\partial X_\mu} \phi + \left( \frac{\partial}{\partial X_\mu} \phi \right)^+ \gamma_\mu^+ \phi = 0 \tag{4.2}$$

Consider the conditions of Eqs. (2.5) and (2.9), and simplify Eq. (4.2) to obtain the differential equation of probability conservation

$$\frac{\partial}{\partial X_\mu} J_\mu = 0 \tag{4.3}$$

or denote as

$$\bar{\nabla} \cdot \bar{J} + \frac{\partial}{\partial t} \rho = 0 \tag{4.4}$$

Four-dimensional probability current density can be written as

$$J_\mu = \rho + j\bar{J} = \phi^+ \gamma_\mu \phi \tag{4.5}$$

where three-dimensional probability current density $J_i (i = 1,2,3)$ and probability density $\rho$ can be written as, respectively

$$J_i = \phi^+ j \gamma_i \phi, \quad \rho = \phi^+ \gamma_4 \phi = \phi^+ \phi \tag{4.6}$$

And with property

$$J_i^+ = J_i, \quad \rho^+ = \rho \tag{4.7}$$

Choose one electron in electromagnetic field

$$\begin{cases} A_\mu = \varphi + j\bar{A} & (4.8a) \\ A_\mu^+ = \varphi - j\bar{A} & (4.8b) \end{cases}$$

Substitute Eqs. (4.8) into (2.14), the electron wave equation will be

$$\left[ \gamma_\mu \left( j\hbar \frac{\partial}{\partial X_\mu} + eA_\mu \right) - m_0 c \right] \phi = 0 \tag{4.9}$$

Taking Hermitian conjugate of Eq. (4.9), we get

$$\left[\gamma_\mu^+ \left(j\hbar(\frac{\partial}{\partial X_\mu})^+ - eA_\mu^+\right) + m_0 c\right]\phi^+ = 0 \tag{4.10}$$

From Eqs. (2.5) and (4.8), we obtain

$$\gamma_\mu^+ A_\mu^+ = \gamma_\mu A_\mu \tag{4.11}$$

Thus Eq. (4.10) can be changed into

$$\left[\gamma_\mu \left(j\hbar \frac{\partial}{\partial X_\mu} - eA_\mu\right) + m_0 c\right]\phi^+ = 0 \tag{4.12}$$

From Eqs. (3.13) and (3.15), let $G = \varepsilon_{\mu\gamma}$, choose

$$\begin{cases} \phi^G = \phi^+ & (4.13a) \\ -\gamma_\mu = \gamma_\mu^G = G\gamma_\mu G & (4.13b) \end{cases}$$

Substitute Eqs. (4.13) into (4.12) as follows

$$\gamma_\mu^G \left(j\hbar \frac{\partial}{\partial X_\mu} - eA_\mu\right)\phi^G - m_0 c \phi^G = 0 \tag{4.14}$$

From Eqs.(2.6) and (4.14), it can be seen that $A_\mu$ is a function of $X_\mu$, and the inversion of $X_\mu$ is equal to that of $A_\mu$. Eq. (4.14) corresponds to the joint inversion of time, space and electric charge in Eq. (4.9) (CPT inversion). It can be considered that Eq. (4.14) corresponds to the wave equation of antiparticles corresponding to Eq. (4.9), whose particles and antiparticles are Hermitian conjugate each other. So probability interpretation $\phi^+ \phi$ of wave function is determined by both particle and antiparticle. Hypothesis of $\phi^+ \phi$ is probability of discovering particles that may be particles or antiparticles.

## 5. Geometric Correlation of Dirac Particles and Minkowski Spaces

From Eq. (1.8), Dirac equations (2.14) and (2.15) can be written as

$$\gamma_\mu \hat{P}_\mu^+ \phi = m_0 c \phi \tag{5.1}$$

$$\phi^+ \hat{P}_\mu \gamma_\mu^+ = m_0 c \phi^+ \tag{5.2}$$

Take spinor wave function

$$\phi = \omega e^{j(\vec{r}\cdot\vec{p}-Et)/\hbar}, \quad \phi^+ = \omega^+ e^{-j(\vec{r}\cdot\vec{p}-Et)/\hbar} \tag{5.3}$$

substitute Eqs. (1.8) and (5.3) into Eqs. (5.1) and (5.2). Then

$$\gamma_\mu \hat{P}_\mu^+ \phi = \gamma_\mu P_\mu^+ \phi = m_0 c \phi \tag{5.4}$$

$$\phi^+ \hat{P}_\mu \gamma_\mu^+ = \phi^+ P_\mu \gamma_\mu^+ = m_0 c \phi^+ \tag{5.5}$$

According to Eqs.(1.5) and (1,7), energy and momentum equations of Eqs. (5.4) and (5.5) can be written as, respectively

$$\begin{cases} \gamma_0 \hat{E}\phi = \gamma_0 E\phi & (5.6a) \\ \gamma_0^+ \hat{E}\phi^+ = \gamma_0^+ E\phi^+ & (5.6b) \end{cases}$$

$$\begin{cases} \gamma_i \hat{p}_i \phi = \gamma_i p \phi \quad (i=1,2,3) & (5.7a) \\ \gamma_i^+ \hat{p}_i \phi^+ = \gamma_i^+ p_i \phi^+ \quad (i=1,2,3) & (5.7b) \end{cases}$$

where Eqs. (5.6a) and (5.7a) can be regarded as energy eigenequation and momentum eigenequation of particles, respectively. And Eqs. (5.6b) and (5.7b) look as those of antiparticles, respectively. Meanwhile, Eqs. (5.1) and (5.2) are regarded as mass eigenequation of particles and antiparticles. It can be proved that rest mass $m_0$ is invariant under coordinate transformation in hyperbolic Minkowski space.

Let two hyperbolic Minkowski momentum spaces $H(E, j\vec{p})$ and $H'(E', j\vec{p}')$ move mutually at velocity $\vec{v}\ (v_x, v_y, v_z)$. General Lorentz transformation formula of four-dimensional momentum can be written as (5)

$$\begin{cases} \vec{p}' = \vec{p} + (\frac{1}{\alpha}-1)\frac{(\vec{p}\cdot\vec{v})\vec{v}}{v^2} + \frac{\vec{v}E}{\alpha c^2} \\ E' = \frac{1}{\alpha}(E+\vec{p}\cdot\vec{v}) \end{cases} \tag{5.8}$$

Eq. (5.8) can be defined by four-dimensional matrix such that

$$P'_\mu = U_{\mu\nu} P_\nu \tag{5.9}$$

which has the same form with Eq. (3.1). Expand Eq. (5.9) as

$$\begin{pmatrix} jp'_x \\ jp'_y \\ jp'_z \\ E' \end{pmatrix} = \begin{pmatrix} 1+(\frac{1}{\alpha}-1)\frac{v_x^2}{v^2} & (\frac{1}{\alpha}-1)\frac{v_x v_y}{v^2} & (\frac{1}{\alpha}-1)\frac{v_x v_z}{v^2} & \frac{jv_x}{\alpha} \\ (\frac{1}{\alpha}-1)\frac{v_x v_y}{v^2} & 1+(\frac{1}{\alpha}-1)\frac{v_y^2}{v^2} & (\frac{1}{\alpha}-1)\frac{v_y v_z}{v^2} & \frac{jv_y}{\alpha} \\ (\frac{1}{\alpha}-1)\frac{v_x v_z}{v^2} & (\frac{1}{\alpha}-1)\frac{v_y v_z}{v^2} & 1+(\frac{1}{\alpha}-1)\frac{v_z^2}{v^2} & \frac{jv_z}{\alpha} \\ \frac{jv_x}{\alpha} & \frac{jv_y}{\alpha} & \frac{jv_z}{\alpha} & \frac{1}{\alpha} \end{pmatrix} \begin{pmatrix} jp_x \\ jp_y \\ jp_z \\ E \end{pmatrix} \quad (5.10)$$

Take inner product of Eq. (5.10), which fulfills Eq. (3.2). We get the relations

$$U^+_{\mu\nu} U_{\mu\nu} = U_{\mu\nu} U^+_{\mu\nu} = I \quad (5.11)$$

The invariant of momentum space interval with a similar form to Eq.(1.6) is

$$P'^+_\mu P'_\mu = P^+_\mu P_\mu = m_0^2 c^2 \quad (5.12)$$

where $m_0^2$ is invariant under coordinate transformation in the four-dimensional momentum space, which is also mass eigenvalue of Dirac equations (5.1) and (5.2).

It is analyzed that the particles and antiparticles are Hermitian conjugate mutually and CPT inversion in above paragraphs. When $e = 0$, Hermitian conjugate of uncharged particle and antiparticle is equivalent to space-time inversion. Space-time inversion of Eqs. (2.14) and (2.15) ($t \to -t, \vec{r} \to -\vec{r}$) corresponds to mass inversion ($jm \to -jm$, or $\hat{m} \to \hat{m}^+$). $\hat{m}$ and $\hat{m}^+$ are defined as mass operators of the particles and antiparticles, respectively. In the meantime, we can find the space-time corresponding points qualitatively in hyperbolic Minkowski space, which are located in timelike region with mutual negative elements ($C_i$ and $-C_i$) as shown in Fig. 1. For instance, take complex quaternion or space-time point $H(X_\mu) = H(t, j\vec{r})$ in $C_1$ region of the first quadrant in the four-dimensional hyperbolic momentum space. Then its space-time inversion point $H(-X_\mu) = H(-t, -j\vec{r})$ is located in $C_5$ region of the third quadrant.

## 6. Feynman Diagram in Minkowski Space

In general, the direct sum of complex numbers in the first quadrant ($C_1$ region) and the third quadrant ($C_5$ region) can be treated as the subtraction of two complex numbers in a same quadrant (such as $C_1$ region). In hyperbolic Minkowski space, the subtraction of hyperbolic complex numbers $X_n - X_m = X_l$ means that the real part and imaginary part subtract respectively and then add. While $X_l$ belongs to timelike region, lightlike region or spacelike region, respectively, according to different conditions, which is the particular characteristics of hyperbolic Minkowski space. Moreover, hyperbolic Minkowski space is characteristic of space-time anisotropy, which makes the timelike or spacelike region $C$ formally correlate to zero factor in lightlike region $\Xi$. Subtraction of element numbers in a same quadrant (or the direct sum of element numbers in $C_i$ and $-C_i$) in hyperbolic Minkowski space takes on the following properties (*6*).

**Property 1** In the lightlike region $\Xi$, if $\theta_n$, $\theta_m \in \Xi$ and they are the same kind zero factors. Then $\theta_n - \theta_m \in \Xi$ is also the same kind zero factor.

**Property 2** In the same timelike or spacelike region ($C_i$), if $X_n$, $X_m \in C_i$, and the straight line through the two points $X_n$ and $X_m$ (called world line) is parallel or vertical to $\Xi$. Then

$$X_n - X_m = a\theta \in \Xi \qquad (6.1)$$

**Property 3** In the same timelike or spacelike region ($C_i$), if $X_n$, $X_m \in C_i$, and the straight line through the two points $X_n$ and $X_m$ is neither parallel nor vertical to $\Xi$. Then

$$X_n - X_m = X_l \in C \qquad (6.2)$$

Take $X_l = ct + jr$, when $ct > r$, $X_l$ is a space-time point in timelike region; when $ct < r$, $X_l$ is the one in spacelike region.

For example, In the first quadrant $C_1$ in timelike region of two-dimensional hyperbolic Minkowski space, let space-time points $X_1 = 8 + j$, $X_2 = 10 + 3j \in C_1$.

Then the line connecting $X_1$ and $X_2$ is parallel to lightlike region $\Xi_1$ (as shown in Fig. 1). The formula $\Delta X = (X_2 - X_1) = 2 + 2j \in \Xi_1$ fulfills $\Delta X^* \Delta X = 0$. Then take $X_3 = 12 + j \in C_1$, the line connecting $X_3$ and $X_2$ is parallel to lightlike region $\Xi_4$ or (perpendicular to $\Xi_1$). The relation is $\Delta X^* = (X_3 - X_2) = 2 - 2j \in \Xi_4$, which also satisfies $\Delta X^* \Delta X = 0$.

In timelike region of hyperbolic Minkowski space, lightlike interval of two physics events is that the two micro objects with rest mass generate a causal relationship through optical signal. That is, when the two physics events interchange photons mass and energy mutually transform. In four-dimensional hyperbolic momentum space $H(E, j\vec{p})$, take four-dimensional momentum $P_\mu, P_\nu$ from Eq. (1.5). For the directional singularity in the Minkowski complex space, when $P_\mu$ and $P_\nu$ are space-time points in the same timelike region of $H(E, j\vec{p})$, the line connecting $P_\mu$ and $P_\nu$ is parallel or vertical to lightlike region $\Xi$. According to Eq. (6.1), the formula is

$$P_\mu - P_\nu = b\vartheta \in \Xi \tag{6.3}$$

where $P_\mu$ and $-P_\nu$ can be seen as the direct sum of the Dirac particles and antiparticles corresponding to Eqs. (5.1) and (5.2). $b\vartheta$ is related to photons (7). Equation (6.3) denotes the coupling relation between the particles /antiparticles with rest mass and the photons in hyperbolic Minkowski space, which provides a geometric interpretation for generation or annihilation of the particles and antiparticles, and the coupling between particles with rest mass and photons in Minkowski space.

Figure 2 introduces Feynman diagram in four-dimensional Minkowski space. If $X_\mu$ represents space-time point of particles $\hat{m}_\mu = jm_{\mu 0}$, $-X_\nu$ is that of antiparticles $\hat{m}_\nu^* = -jm_{\nu 0}$. While $a\theta$ is a space-time point of photons, which locates in lightlike region $\Xi$. Figure 2(a) and 2(b) show Feynman diagram that the particles and antiparticles coupling with virtual photon.

## 7. Klein-Gordon Equation and Schrödinger Equation

Apply Eq. (1.6) to the state function $\Phi(\vec{r},t)$

$$\hat{P}_\mu^+ \hat{P}_\mu \Phi = m_0^2 c^2 \Phi \tag{7.1}$$

Equation (1.12) can be written as Klein-Gordon equation

$$(\hbar^2 c^2 \nabla^2 - \hbar^2 \frac{\partial^2}{\partial t^2} - m_0^2 c^4)\Phi = 0 \tag{7.2}$$

or

$$\frac{1}{c^2}\frac{\partial^2}{\partial t^2}\Phi - \nabla^2\Phi + \frac{m_0^2 c^2}{\hbar^2}\Phi = 0 \tag{7.3}$$

For Eq. (7.3) is the compound equation of particle Dirac wave equation Eq. (2.14) and antiparticle Dirac wave equation Eq. (2.15), it is likely that a Klein-Gordon particle compounds by the spinor particle with spin ½ and the spinor antiparticle with spin -½, and $\Phi = \phi^+\phi$ is the state function of Klein-Gordon equation. Due to Dirac spinor state function has the matrix form with four components, and its inner product $\phi^+\phi$ has the real form with one component, so Klein-Gordon equation is a real equation with spin zero.

Making $\Phi^+ \times (7.2) - \Phi \times (7.2)^+$, we have

$$\hbar^2 c^2 \nabla(\Phi^+\nabla\Phi - \Phi\nabla\Phi^+) = \hbar^2 \frac{\partial}{\partial t}(\Phi^+\frac{\partial}{\partial t}\Phi - \Phi\frac{\partial}{\partial t}\Phi^+) \tag{7.4}$$

Take probability density and probability current density be

$$\begin{cases} \vec{J} = -j\hbar(\Phi^+\nabla\Phi - \Phi\nabla\Phi^+) \\ \rho = \frac{j\hbar}{c^2}(\Phi^+\frac{\partial}{\partial t}\Phi - \Phi\frac{\partial}{\partial t}\Phi^+) \end{cases} \tag{7.5}$$

which satisfy the conservative relation

$$\frac{\partial}{\partial t}\rho + \nabla \cdot \vec{J} = 0 \tag{7.6}$$

The spin of $\pi$ meson which Klein-Gordon equation corresponds is zero, so the probability density in Eq. (7.5) may be positive, or negative, and also can be zero, which is the inevitable result for compound particles when the particle and antiparticle couple.

In the condition of non-relativistic limit, it exists the relations $\frac{v}{c} << 1$ and $m \approx m_0$. The particle energy satisfies

$$E = m_0 c^2 + \frac{p^2}{2m} \tag{7.7}$$

In Eq. (7.7), the first term is the rest energy of particles, and the second term is the kinetic energy. Equation (7.7) is written as operator form

$$\hat{E} = m_0 c^2 + \frac{\hat{p}^+ \hat{p}}{2m} \tag{7.8}$$

From Eq. (1.7), we have

$$-j\hbar \frac{\partial}{\partial t} = m_0 c^2 - \frac{\hbar^2}{2m} \nabla^2 \tag{7.9}$$

Let state function be

$$\begin{cases} \Phi(\vec{r},t) = e^{\frac{j}{\hbar}(\vec{p}\cdot\vec{r}-Et)} = \varphi(\vec{r},t) e^{-\frac{j}{\hbar}mc^2 t} \\ \varphi(\vec{r},t) = e^{\frac{j}{\hbar}(\vec{p}\cdot\vec{r}-E_k t)} \end{cases} \tag{7.10}$$

Then acting energy operator $\hat{E} = -j\hbar \frac{\partial}{\partial t}$ on the first formula of Eq. (7.10) yields

$$\hat{E}\Phi = -j\hbar \frac{\partial \Phi}{\partial t} = (-j\hbar \frac{\partial \varphi}{\partial t} + m_0 c^2 \varphi) e^{-\frac{j}{\hbar}mc^2 t} \tag{7.11}$$

Combining Eqs. (7.9) and (7.11), we obtain

$$-j\hbar \frac{\partial \Phi}{\partial t} = (-j\hbar \frac{\partial \varphi}{\partial t} + m_0 c^2 \varphi) e^{-\frac{j}{\hbar}mc^2 t} = (m_0 c^2 \varphi - \frac{\hbar^2}{2m} \nabla^2 \varphi) e^{-\frac{j}{\hbar}mc^2 t} \tag{7.12}$$

Simplifying Eq. (7.12) to obtain

$$j\hbar \frac{\partial \varphi}{\partial t} = \frac{\hbar^2}{2m} \nabla^2 \varphi \tag{7.13}$$

Equation (7.13) is Schrödinger equation of free particles in Minkowski space (*8*).

## 8. Action Equation and Mass Renormalization of Dirac Particles

Take the equation of least action in Minkowski space be $S = -\int b\, ds$, where $b = m_0 c$, that is

$$S = -\int m_0 c\, ds \tag{8.1}$$

$ds$ is invariant, referring Eq. (1.2) we take

$$ds = -\sqrt{dX_\mu^+ dX_\mu} \tag{8.2}$$

From Eqs. (1.3) to (1.6), equation (8.1) can be rewritten as

$$S = \int m_0 c \sqrt{dX_\mu^+ dX_\mu} = \int c \sqrt{\frac{m_0 dX_\mu^+}{d\tau} \frac{m_0 dX_\mu}{d\tau}} d\tau = \int c \sqrt{P_\mu^+ P_\mu} d\tau = \int L d\tau \quad (8.3)$$

Lagrangian function is

$$L = E_0 = m_0 c^2 = c \sqrt{P_\mu^+ P_\mu} \quad (8.4)$$

where $E_0$ is rest energy. The operator representation of Eq. (8.4) is

$$\hat{H}_0 = \hat{L} = c \sqrt{\hat{P}_\mu^+ \hat{P}_\mu} \quad (8.5)$$

Equation (8.5) is Lagrangian operator when Dirac particle and antiparticle couple. Equation (8.3) is the integral to rest mass or rest energy

$$S = \int c \sqrt{P_\mu^+ P_\mu} d\tau = \int m_0 c^2 d\tau = \int E_0 d\tau \quad (8.6)$$

For $L = m_0 c^2$ is invariant, taking variation to Eq. (8.6), we obtain

$$\delta S = \int \delta L d\tau = \int c^2 \delta m_0 d\tau = \int \delta E_0 d\tau \quad (8.7)$$

or

$$\delta L = \delta m_0 c^2 = c \delta(P_\mu^+ P_\mu) / 2\sqrt{P_\mu^+ P_\mu} = \frac{1}{2m_0} \delta(P_\mu^+ P_\mu) \quad (8.8)$$

Substituting Eq. (8.8) to Eq. (8.7) yields

$$\delta S = \int \delta L d\tau = \int c^2 \delta m_0 d\tau = \int \frac{1}{2m_0} \delta(P_\mu^+ P_\mu) d\tau \quad (8.9)$$

Generally the integral to rest mass or rest energy in Eq. (8.6) over the whole space is divergent. Consider Eq. (6.3), let the integral path be parallel or vertical to lightlike region in order to get geometric correlation between timelike region and lightlike region, then it is likely to solve the problem that the integral to mass is infinite. In momentum space $H(p_0, j\bar{p}) = H(\frac{E}{c}, j\bar{p})$ which possesses four-dimensional discrete phase lattices, take the straight line through phase lattices $\mu$ and $\nu$ in the same timelike region be parallel or vertical to lightlike region, which are corresponding to four dimensional momentum $P_\mu = m_{\mu 0} \varpi_\mu$ and $P_\nu = m_{\nu 0} \varpi_\nu$, respectively. The subscripts $\mu$ and $\nu$ have dual properties, one denotes different time-space lattices; the other denotes four components of four-dimensional complex element. Dirac equation (2.14) of which is written as, respectively

$$\begin{cases} \gamma_\mu P_\mu^+ \phi_\mu = j\hbar \gamma_\mu \dfrac{\partial}{\partial X_\mu} \phi_\mu = m_{\mu 0} c \phi_\mu \\ \gamma_\nu P_\nu^+ \phi_\nu = j\hbar \gamma_\nu \dfrac{\partial}{\partial X_\nu} \phi_\nu = m_{\nu 0} c \phi_\nu \end{cases} \quad (8.10)$$

Dirac equation (8.10) is can also be denoted as mass eigenequation

$$\begin{cases} \hat{m}_{\mu 0} \phi_\mu = m_{\mu 0} \phi_\mu \\ \hat{m}_{\nu 0} \phi_\nu = m_{\nu 0} c \phi_\nu \end{cases} \quad (8.11)$$

where $\hat{m}_{\mu 0}$ and $\hat{m}_{\nu 0}$ denote four-dimensional mass operators with quarternary form, which satisfy the following relations

$$\begin{cases} \hat{m}_{\mu 0} c = \gamma_\mu \hat{P}_\mu = j\hbar \gamma_\mu \dfrac{\partial}{\partial X_\mu} \\ \hat{m}_{\mu 0}^+ c = \gamma_\mu^+ \hat{P}_\mu^+ = -j\hbar \gamma_\mu^+ \dfrac{\partial}{\partial X_\mu^+} \end{cases} \quad (8.12)$$

Taking the differences of Eq. (8.11), we then have

$$\hat{m}_{\mu 0} \phi_\mu - \hat{m}_{\nu 0} \phi_\nu = m_{\mu 0} \phi_\mu - m_{\nu 0} \phi_\nu \quad (8.13)$$

From Eq. (6.3), when the discrete phase lattices $\mu$ and $\nu$ get causal relation through photon, time variant is equal to space variant (*3*), take

$$j(kx_1 - \varpi t_1) = j[k(x_2 - \Delta x) - \varpi(t_2 - \Delta t)] = j(kx_2 - \varpi t_2) - j(k\Delta x - \varpi \Delta t)$$

If let $a^* = e^{-jk\Delta x}$, $a = e^{j\varpi \Delta t}$, and

$$k\Delta x = \varpi \Delta t, \quad a^* a = 1 \quad (8.14)$$

then the wave function satisfies the following relation

$$\phi_\mu(x_1, t_1) = e^{j(kx_1 - \varpi t_1)} = a^* a \phi(x_2, t_2) = e^{j(kx_2 - \varpi t_2)} = \phi_\nu(x_2, t_2) \quad (8.15)$$

that is, $\phi_\mu$ is equal to $\phi_\nu$, hence Eq. (8.11) can be rewritten as

$$(\hat{m}_{\mu 0} - \hat{m}_{\nu 0}) \phi = (m_{\mu 0} - m_{\nu 0}) \phi \quad (8.16)$$

Let

$$\delta m_0 = \dfrac{m_{\mu 0} - m_{\nu 0}}{2}, \quad m = \dfrac{m_{\mu 0} + m_{\nu 0}}{2} \quad (8.17)$$

where $\delta m_0$ is named electromagnetic mass, and $m$ is natural mass. From Eq. (8.17) mass renormalization relation is written as

$$\begin{cases} m_{\mu 0} = m + \delta m_0 \\ m_{\nu 0} = m - \delta m_0 \end{cases} \quad (8.18)$$

where $m_{\mu 0}$ and $m_{\nu 0}$ are corresponding to the rest mass of time space phase lattices $\mu$ and $\nu$ in four-dimensional Minkowski momentum space, respectively (Fig. 2).

From Eqs. (6.3), (8.12) and (8.16), let

$$\delta \hat{m}_0 = \frac{1}{2}(\hat{m}_{\mu 0} - \hat{m}_{\nu 0}) = \frac{1}{2}(\gamma_\mu \hat{P}_\mu - \gamma_\nu \hat{P}_\nu) = \frac{1}{2}\gamma_\mu \delta \hat{P}_\mu = \frac{1}{2}\gamma_\mu (\frac{\delta \hat{E}}{c} - j\delta \hat{p}) \quad (8.19)$$

and the operators satisfy the following relations

$$[\delta \hat{E}, \delta \hat{p}] = \delta \hat{E} \delta \hat{p} - \delta \hat{p} \delta \hat{E} = 0, \quad (\frac{\delta \hat{E}}{c})^2 - (\delta \hat{p})^2 = 0$$

Then we have

$$\delta \hat{m}_0^+ \delta \hat{m}_0 = \frac{1}{4}[(\hat{m}_{\mu 0} - \hat{m}_{\nu 0})^+ (\hat{m}_{\mu 0} - \hat{m}_{\nu 0})] = 0 \quad (8.20)$$

$\delta \hat{m}_0$ and $\delta \hat{m}_0^+$ are mutually orthogonal in Minkowski lightlike region. Then the operators of timelike region and lighlike region get relation by Eq. (8.20). From the first part of Eq. (8.17), we take

$$2\delta m_0 c^2 = m_{\mu 0} c^2 - m_{\nu 0} c^2 = 2\hbar \varpi \quad (8.21)$$

where $\hbar \varpi$ is the energy of photon, equation (8.21) shows the energy conservation relation when the particle with rest mass couples with the photon in space-time phase lattices $\mu$ and $\nu$ which parallel or vertical to lightlike region. Expending Eq. (8.20) we obtain

$$\hat{m}_{\mu 0}^+ \hat{m}_{\nu 0} + \hat{m}_{\mu 0}^+ \hat{m}_{\nu 0} = (\hat{m}_{\mu 0})^2 + (\hat{m}_{\nu 0})^2 \quad (8.22)$$

Equation (7.1) can also be written as

$$\gamma_\nu^+ \gamma_\mu \hat{P}_\nu^+ \hat{P}_\mu \Phi = \hat{P}_\nu^+ \hat{P}_\mu \Phi = c^2 \hat{m}_{\nu 0} \hat{m}_{\mu 0}^+ \Phi = c^2 m_{\nu 0} m_{\mu 0} \Phi \quad (8.23)$$

or

$$\gamma_\mu^+ \gamma_\nu \hat{P}_\mu^+ \hat{P}_\nu \Phi = \hat{P}_\mu^+ \hat{P}_\nu \Phi = c^2 \hat{m}_{\mu 0} \hat{m}_{\nu 0}^+ \Phi = c^2 m_{\mu 0} m_{\nu 0} \Phi \quad (8.24)$$

Equations (8.23) and (8.24) are of equivalence.

## 9. Geometric Correlation with Yang-Mills Equation

In electromagnetic field rest mass operators $\hat{m}_{\mu 0}$ and $\hat{m}_{\nu 0}$ can be written as, respectively

$$\begin{cases} \hat{m}_{\mu 0} = j\gamma_\mu \dfrac{1}{c}(\hbar\dfrac{\partial}{\partial X_\mu} + jeA_\mu) \\ \hat{m}_{\nu 0} = j\gamma_\nu \dfrac{1}{c}(\hbar\dfrac{\partial}{\partial X_\nu} + jeA_\nu) \end{cases} \quad (9.1)$$

And the corresponding Hermitian conjugates are

$$\begin{cases} \hat{m}_{\mu 0}^+ = -j\gamma_\mu^+ \dfrac{1}{c}(\hbar\dfrac{\partial}{\partial X_\mu^+} - jeA_\mu^+) \\ \hat{m}_{\nu 0}^+ = -j\gamma_\nu^+ \dfrac{1}{c}(\hbar\dfrac{\partial}{\partial X_\nu^+} - jeA_\nu^+) \end{cases} \quad (9.2)$$

From Eqs. (1.8), (8.6), (8.23) and (8.24), making Eq. (9.1) subtract Eq. (9.2), then integrating in electromagnetic field, we get

$$0 = \int_\mu^\nu (\hat{m}_{\mu 0}^+ \phi^+ \hat{m}_{\nu 0} \phi - \hat{m}_{\nu 0}^+ \phi^+ \hat{m}_{\mu 0} \phi) c^2 d\tau$$

$$= \int_\mu^\nu [(\hat{P}_\mu + eA_\mu^+)\phi^+ \gamma_\mu^+ \gamma_\nu (\hat{P}_\nu^+ + eA_\nu)\phi - (\hat{P}_\nu + eA_\nu^+)\phi^+ \gamma_\nu^+ \gamma_\mu (\hat{P}_\mu^+ + eA_\mu)\phi] d\tau$$

$$= \int_\mu^\nu [(-j\hbar\dfrac{\partial}{\partial X_\mu^+} + eA_\mu^+)\phi^+ (j\hbar\dfrac{\partial}{\partial X_\nu} + eA_\nu)\phi - (-j\hbar\dfrac{\partial}{\partial X_\nu^+} + eA_\nu^+)\phi^+ (j\hbar\dfrac{\partial}{\partial X_\mu} + eA_\mu)\phi] d\tau$$

$$= \int_\mu^\nu [\hbar^2(\dfrac{\partial \phi^+}{\partial X_\nu^+}\dfrac{\partial \phi}{\partial X_\mu} - \dfrac{\partial \phi^+}{\partial X_\mu^+}\dfrac{\partial \phi}{\partial X_\nu}) + je\hbar(\dfrac{\partial \phi^+}{\partial X_\nu^+}A_\mu \phi - \dfrac{\partial \phi^+}{\partial X_\mu^+}A_\nu \phi)$$

$$- e^2(A_\nu^+ A_\mu - A_\mu^+ A_\nu)\phi^+ \phi) - je\hbar(A_\nu^+ \phi^+ \dfrac{\partial \phi}{\partial X_\mu} - A_\mu^+ \phi^+ \dfrac{\partial \phi}{\partial X_\nu})] d\tau \quad (9.3)$$

Equation (9.3) is the related expression for the four Dirac wave equations of particles and antiparticles (9). The differential of composite functions satisfies

$$\dfrac{\partial(A_\nu \phi^+ \phi)}{\partial X_\mu^+} = A_\nu \phi^+ \dfrac{\partial \phi}{\partial X_\mu^+} + A_\nu \phi \dfrac{\partial \phi^+}{\partial X_\mu^+} + \phi^+ \phi \dfrac{\partial A_\nu}{\partial X_\mu^+} \quad (9.4)$$

then

$$A_\nu \phi^+ \dfrac{\partial \phi}{\partial X_\mu^+} + \dfrac{\partial \phi^+}{\partial X_\mu^+} A_\nu \phi = \dfrac{\partial(A_\nu \phi^+ \phi)}{\partial X_\mu^+} - \phi^+ \phi \dfrac{\partial A_\nu}{\partial X_\mu^+} \quad (9.5)$$

From Eqs. (1.9) and (4.1), and notice the relation $\dfrac{\partial \phi}{\partial X_\mu} A_\nu^+ = \dfrac{\partial \phi}{\partial X_\mu^+} A_\nu$, then Eq. (9.3) can be written as

$$\int_\mu^\nu [(\hat{P}_\mu + eA_\mu^+)\phi^+ (\hat{P}_\nu^+ + eA_\nu)\phi - (\hat{P}_\nu + eA_\nu^+)\phi^+ (\hat{P}_\mu^+ + eA_\mu)\phi] d\tau$$

$$= \int_\mu^\nu [\hbar^2(\frac{\partial \phi^+}{\partial X_\nu^+}\frac{\partial \phi}{\partial X_\mu} - \frac{\partial \phi^+}{\partial X_\mu^+}\frac{\partial \phi}{\partial X_\nu}) + je\hbar(\frac{\partial(A_\mu \phi^+ \phi)}{\partial X_\nu^+} - \frac{\partial(A_\nu \phi^+ \phi)}{\partial X_\mu^+})]d\tau$$

$$-\int_\mu^\nu [je\hbar(\frac{\partial A_\mu}{\partial X_\nu^+} - \frac{\partial A_\nu}{\partial X_\mu^+}) + e^2(A_\nu^+ A_\mu - A_\mu^+ A_\nu)]\phi^+ \phi d\tau = 0 \qquad (9.6)$$

Writing Dirac spinor equation of particles and antiparticles Eq. (4.9) as the electromagnetic equation of phase lattice $\mu$, multiplying $\phi^+ \gamma_\mu^+$, and taking partial derivative with respect to $X_\nu^+$, we obtain

$$j\hbar \frac{\partial}{\partial X_\nu^+}(\phi^+ \frac{\partial \phi}{\partial X_\mu}) + e\frac{\partial(A_\mu \phi^+ \phi)}{\partial X_\nu^+} = c\frac{\partial(m_{\mu 0}\phi^+ \gamma_\mu^+ \phi)}{\partial X_\nu^+} = \frac{1}{c}\frac{\partial(E_{\mu 0}J_\mu^+)}{\partial X_\nu^+} \qquad (9.7)$$

As the same, writing Eq. (4.9) as the electromagnetic equation of phase lattice $\nu$, multiplying $\phi^+ \gamma_\nu^+$, and taking partial derivative with respect to $X_\mu^+$, we obtain

$$j\hbar \frac{\partial}{\partial X_\mu^+}(\phi^+ \frac{\partial \phi}{\partial X_\nu}) + e\frac{\partial(A_\nu \phi^+ \phi)}{\partial X_\mu^+} = c\frac{\partial(m_\nu \phi^+ \gamma_\nu^+ \phi)}{\partial X_\mu^+} = \frac{1}{c}\frac{\partial(E_{\nu 0}J_\nu^+)}{\partial X_\mu^+} \qquad (9.8)$$

Here four-dimensional probability current density $J_\mu = \rho + j\vec{J} = \phi^+ \gamma_\mu \phi$, which satisfies Eq. (4.5). Making the difference between Eqs. (9.7) and (9.8) yields

$$\frac{\partial(A_\mu \phi^+ \phi)}{\partial X_\nu^+} - \frac{\partial(A_\nu \phi^+ \phi)}{\partial X_\mu^+}$$

$$= -\frac{j\hbar}{e}[\frac{\partial}{\partial X_\nu^+}(\phi^+ \frac{\partial \phi}{\partial X_\mu}) - \frac{\partial}{\partial X_\mu^+}(\phi^+ \frac{\partial \phi}{\partial X_\nu})] - \frac{1}{ce}(\frac{\partial E_{\nu 0}J_\nu^+}{\partial X_\mu^+} - \frac{\partial E_{\mu 0}J_\mu^+}{\partial X_\nu^+})$$

$$= -\frac{j\hbar}{e}[(\frac{\partial \phi^+}{\partial X_\nu^+}\frac{\partial \phi}{\partial X_\mu} - \frac{\partial \phi^+}{\partial X_\mu^+}\frac{\partial \phi}{\partial X_\nu}) - (\phi^+ \frac{\partial^2 \phi}{\partial X_\nu^+ \partial X_\mu} - \phi^+ \frac{\partial^2 \phi}{\partial X_\mu^+ \partial X_\nu})]$$

$$-\frac{1}{ce}(\frac{\partial E_{\nu 0}J_\nu^+}{\partial X_\mu^+} - \frac{\partial E_{\mu 0}J_\mu^+}{\partial X_\nu^+}) = -\frac{j\hbar}{e}(\frac{\partial \phi^+}{\partial X_\nu^+}\frac{\partial \phi}{\partial X_\mu} - \frac{\partial \phi^+}{\partial X_\mu^+}\frac{\partial \phi}{\partial X_\nu}) - \frac{1}{ce}(J_\nu^+ \frac{\partial E_{\nu 0}}{\partial X_\mu^+} - J_\mu^+ \frac{\partial E_{\mu 0}}{\partial X_\nu^+})$$

(9.9)

where we use the relations $\frac{\partial^2}{\partial X_\mu^+ \partial X_\nu} = \frac{\partial^2}{\partial X_\nu^+ \partial X_\mu}$. Substituting Eq. (9.9) into Eq. (9.6) yields

$$\int_\mu^\nu [e^2(A_\mu^+ A_\nu - A_\nu^+ A_\mu) + je\hbar(\frac{\partial A_\nu}{\partial X_\mu^+} - \frac{\partial A_\mu}{\partial X_\nu^+})]\phi^+ \phi d\tau = -\int_\mu^\nu \frac{j\hbar}{c}(J_\nu^+ \frac{\partial E_{\nu 0}}{\partial X_\mu^+} - J_\mu^+ \frac{\partial E_{\mu 0}}{\partial X_\nu^+})d\tau$$

$$= -\int_\mu^\nu \frac{j\hbar}{c}(\gamma_\nu^+ \frac{\partial E_{\nu 0}}{\partial X_\mu^+} - \gamma_\mu^+ \frac{\partial E_{\mu 0}}{\partial X_\nu^+})]\phi^+\phi d\tau \tag{9.10}$$

or

$$je\hbar(\frac{\partial A_\nu}{\partial X_\mu^+} - \frac{\partial A_\mu}{\partial X_\nu^+}) + e^2(A_\mu^+ A_\nu - A_\nu^+ A_\mu) = -\frac{j\hbar}{c}(\gamma_\nu^+ \frac{\partial E_{\nu 0}}{\partial X_\mu^+} - \gamma_\mu^+ \frac{\partial E_{\mu 0}}{\partial X_\nu^+}) \tag{9.11}$$

Equation (9.11) can be taken as the general Yang-mills equation in hyperbolic Minkowski space. Let

$$F'_{\mu\nu} = je\hbar(\frac{\partial A_\nu}{\partial X_\mu^+} - \frac{\partial A_\mu}{\partial X_\nu^+}) + e^2(A_\mu^+ A_\nu - A_\nu^+ A_\mu) \tag{9.12}$$

Equation (9.12) is the traditional form of Yang-Mills equation (*10*), but Eq. (9.12) hasn't given the transformation relation between electromagnetic mass and energy including in the right part of Eq. (9.11).

In Eq. (9.11) $\frac{\partial E_{\mu 0}}{\partial X_\nu^+}$ and $\frac{\partial E_{\nu 0}}{\partial X_\mu^+}$ are the changes of rest energy when the particles at $P_\mu$ and $P_\nu$ couple with photons respectively. Let (*11*)

$$\begin{cases} \delta E_{\nu 0} = E_{\mu 0} - E_{\nu 0} = \frac{\partial E_{\nu 0}}{\partial X_\mu^+}\delta X_\mu^+ = 2\delta m_0 c^2 = 2\hbar\varpi \\ \delta E_{\mu 0} = E_{\nu 0} - E_{\mu 0} = \frac{\partial E_{\mu 0}}{\partial X_\nu^+}\delta X_\nu^+ = -2\delta m_0 c^2 = -2\hbar\varpi \end{cases} \tag{9.13}$$

Combing Eqs. (9.13) and (9.11), we have

$$F'_{\mu\nu} = je\hbar(\frac{\partial A_\nu}{\partial X_\mu^+} - \frac{\partial A_\mu}{\partial X_\nu^+}) + e^2(A_\mu^+ A_\nu - A_\nu^+ A_\mu) = -j\frac{2\hbar^2\varpi}{c}(\frac{\gamma_\nu^+}{\delta X_\mu^+} + \frac{\gamma_\mu^+}{\delta X_\nu^+}) \tag{9.14}$$

Equation (9.14) is the corresponding Yang-mills equation of excited photons when the particle couples with the antiparticle and of the conversion of electromagnetic mass and energy.

## 10. Geometric Correlation with Maxwell Equation

Take variation to Eq. (9.10), we have

$$\int_\mu^\nu \delta[e^2(A_\mu^+ A_\nu - A_\nu^+ A_\mu) + je\hbar(\frac{\partial A_\nu}{\partial X_\mu^+} - \frac{\partial A_\mu}{\partial X_\nu^+})]\phi^+\phi d\tau = -\int_\mu^\nu \frac{j\hbar}{c}\delta(J_\nu^+ \frac{\partial E_{\nu 0}}{\partial X_\mu^+} - J_\mu^+ \frac{\partial E_{\mu 0}}{\partial X_\nu^+})d\tau$$

(10.1)

By using $\delta$ function for the left side of Eq. (10.1), and canceling the derivative and integral terms for the right side of which, we have

$$\delta[e^2(A_\mu^+ A_\nu - A_\nu^+ A_\mu) + je\hbar(\frac{\partial A_\nu}{\partial X_\mu^+} - \frac{\partial A_\mu}{\partial X_\nu^+})] = -\frac{j\hbar}{c}(J_\nu^+ \frac{\partial E_{\nu 0}}{\partial X_\mu^+} - J_\mu^+ \frac{\partial E_{\mu 0}}{\partial X_\nu^+}) \tag{10.2}$$

Taking variation with respect to $X_\mu$ and $X_\nu$ to the left part of Eq. (10.2), and considering Eq. (9.14) we obtain

$$je\hbar\frac{\partial}{\partial X_\mu}(\frac{\partial A_\nu}{\partial X_\mu^+}-\frac{\partial A_\mu}{\partial X_\nu^+})\delta X_\mu + je\hbar\frac{\partial}{\partial X_\nu}(\frac{\partial A_\nu}{\partial X_\mu^+}-\frac{\partial A_\mu}{\partial X_\nu^+})\delta X_\nu$$

$$+e^2(A_\mu^+\frac{\partial A_\nu}{\partial X_\mu}-\frac{\partial A_\nu^+}{\partial X_\mu}A_\mu)\delta X_\mu + e^2(\frac{\partial A_\mu^+}{\partial X_\nu}A_\nu - A_\nu^+\frac{\partial A_\mu}{\partial X_\nu})\delta X_\nu$$

$$=-j\frac{2\hbar^2\varpi}{c}(\frac{J_\nu^+}{\delta X_\mu^+}+\frac{J_\mu^+}{\delta X_\nu^+}) \tag{10.3}$$

It is likely that the third and fourth terms of Eq. (10.3) are related to strong interactions, which satisfy gauge condition due to the time fluidity of four-dimensional time space,

$$\frac{\partial A_\mu^+}{\partial X_\mu}=\frac{\partial A_\mu}{\partial X_\mu}=\frac{\partial A_\nu^+}{\partial X_\nu}=\frac{\partial A_\nu}{\partial X_\nu}=0 \tag{10.4}$$

Equation (10.3) can be separated as

$$je\hbar\frac{\partial}{\partial X_\mu}(\frac{\partial A_\nu}{\partial X_\mu^+}-\frac{\partial A_\mu}{\partial X_\nu^+})+e^2(A_\mu^+\frac{\partial A_\nu}{\partial X_\mu}-\frac{\partial A_\nu^+}{\partial X_\mu}A_\mu)=-j\frac{2\hbar^2\varpi}{c\delta X_\mu^+\delta X_\mu}J_\nu^+ \tag{10.5}$$

$$je\hbar\frac{\partial}{\partial X_\nu}(\frac{\partial A_\nu}{\partial X_\mu^+}-\frac{\partial A_\mu}{\partial X_\nu^+})+e^2(\frac{\partial A_\mu^+}{\partial X_\nu}A_\nu-A_\nu^+\frac{\partial A_\mu}{\partial X_\nu})=-j\frac{2\hbar^2\varpi}{c\delta X_\nu^+\delta X_\nu}J_\mu^+ \tag{10.6}$$

Equations (10.5) and (10.6) have opposite four-dimensional probability current density. If not considering the second term, Equation (10.5) may be rewritten as

$$\frac{\partial}{\partial X_\mu^+}(\frac{\partial A_\nu}{\partial X_\mu}-\frac{\partial A_\mu}{\partial X_\nu})\delta X_\mu^+\delta X_\mu = \frac{\partial F_{\mu\nu}}{\partial X_\mu^+}\delta X_\mu^+\delta X_\mu = -\frac{2h\nu}{ce}J_\nu^+ \tag{10.7}$$

where antisymmetric matrix tensor $F_{\mu\nu}$ is

$$F_{\mu\nu}=\frac{\partial A_\nu}{\partial X_\mu}-\frac{\partial A_\mu}{\partial X_\nu} \tag{10.8}$$

From Eqs. (1.1) and (1.2), $\delta X_\mu^+\delta X_\mu$ in Eq. (10.7) is taken as the time-space interval between $\mu$ and $\nu$, then let

$$\delta X_\mu^+\delta X_\mu=\frac{\lambda_c^2}{4} \tag{10.9}$$

and substitute the relation to Eq. (10.7) to obtain

$$\frac{\partial F_{\mu\nu}}{\partial X_\mu^+} = -\frac{8h\nu}{ce\lambda_c^2} J_\nu^+ \tag{10.10}$$

If Eq. (9.1) take the form

$$\hat{m}_{\mu 0} = j\gamma_\mu \frac{1}{c}(\hbar \frac{\partial}{\partial X_\mu} + j\frac{1}{e} A_\mu) \tag{10.11}$$

Then Eq. (10.10) can be rewritten as

$$\frac{\partial F_{\mu\nu}}{\partial X_\mu^+} = -\frac{8h\nu e}{c\lambda_c^2} J_\nu^+ = -\frac{8h\nu \varepsilon_0}{\lambda_c^2} ec\mu_0 J_\nu^+ \tag{10.12}$$

Take $h\nu = m_0 c^2$, which is the rest energy of an electron, and $\lambda_c$ is Compton wavelength, then

$$\frac{8h\nu\varepsilon_0}{\lambda_c^2} = \frac{8m_0 c^2 \varepsilon_0}{\lambda_c^2} = \frac{8 \times 9.109 \times 10^{-31} \times 3^2 \times 10^{16} \times 8.854 \times 10^{-12}}{2.426^2 \times 10^{-24}} = 0.987 \approx 1 \tag{10.13}$$

Let

$$ecJ_\nu = I_\nu, \quad I_\nu = j\vec{I} + c\rho \tag{10.14}$$

where $I_\nu$ is four-dimensional current density, $\rho = eJ_4$ is charge density, and $\vec{I} = ec\vec{J}$ is current density. Equation (10.12) is written as

$$\frac{\partial F_{\mu\nu}}{\partial X_\mu^+} = -\mu_0 I_\nu^+ \tag{10.15}$$

Equation (10.15) is the corresponding four-dimensional Maxwell equation when the particle couples with the antiparticle, the left side of which represents the electromagnetic change, and the right side is four-dimensional current density.

Due to $\lambda_c = \frac{h}{m_0 c}$, and noting Eqs. (1.3) and (1.4), we rewrite Eq. (10.9) as

$$\varpi_\mu^+ \varpi_\mu (\delta\tau)^2 = c^2 (\delta\tau)^2 = \frac{\lambda_c^2}{4} = \frac{h^2}{4m_0^2 c^2} \tag{10.16}$$

or

$$\delta E \delta\tau = h \tag{10.17}$$

where $\delta E = 2m_0 c^2$ is the rest energy of the electron. Let

$$\delta\tau = \frac{h}{2m_0 c^2} = 4.041 \times 10^{-21} s \tag{10.18}$$

Equation (10.17) is the quantized condition of excited electromagnetism when the particle couples with the antiparticle, which is related to micro objects $\mu$ and $\nu$ lying different non-particle phase lattices. Equation (10.18) is the time interval of the mass-energy conversion.

Antisymmetric matrix tensor $F_{\mu\nu}$ in Eq. (10.8) satisfies $F_{\mu\nu} = -F_{\nu\mu}$, which is the rotation of four-dimensional vector potential. From Eqs. (1.9) and (4.10), the antisymmetric matrix form of $F_{\mu\nu}$ is written as

$$F_{\mu\nu} = \begin{pmatrix} 0 & B_3 & -B_2 & j\frac{1}{c}E_1 \\ -B_3 & 0 & B_1 & j\frac{1}{c}E_2 \\ B_2 & -B_1 & 0 & j\frac{1}{c}E_3 \\ -j\frac{1}{c}E_1 & -j\frac{1}{c}E_2 & -j\frac{1}{c}E_3 & 0 \end{pmatrix} \quad (10.19)$$

where

$$E_i = -jc\left(\frac{\partial A_i}{\partial X_4} - \frac{\partial A_4}{\partial X_i}\right), i = 1,2,3 \quad (10.20)$$

$$B_i = \frac{\partial A_k}{\partial X_j} - \frac{\partial A_j}{\partial X_k}, i,j,k = 1,2,3 \quad (10.21)$$

Equation (10.15) is the first and fouth equations of Maxwell equations.

$$\begin{cases} \nabla \cdot \vec{D} = \rho \\ \nabla \times \vec{H} - \frac{\partial \vec{D}}{\partial t} = \vec{I} \end{cases} \quad (10.22)$$

其中 $\vec{D} = \varepsilon_0 \vec{E}$, $\vec{B} = \mu_0 \vec{H}$ 。From Eq. (10.8) we obtain

$$\frac{\partial F_{\nu\lambda}}{\partial X_\mu} + \frac{\partial F_{\lambda\mu}}{\partial X_\nu} + \frac{\partial F_{\mu\nu}}{\partial X_\lambda} = 0 \quad (10.23)$$

Equation (10.23) can be separated

$$\begin{cases} \nabla \cdot \vec{B} = 0 \\ \nabla \times \vec{E} + \frac{1}{c}\frac{\partial \vec{B}}{\partial t} = 0 \end{cases} \quad (10.24)$$

Equation (10.24) is the second and third equations of Maxwell equations.

Now we prove Eq. (10.22) form Eq. (10.15).

When the subscript $\nu = 1$ in Eq. (10.15), from Eq. (4.8), considering Eqs. (10.19), (10.20) and (10.21), we have

$$j\mu_0 I_1 = \frac{\partial}{\partial X_\mu^+} F_{\mu 1}$$

From Eq. (1.9), expanding the above equation

$$j\mu_0 I_1 = -j(\frac{\partial F_{21}}{\partial X_2} + \frac{\partial F_{31}}{\partial X_3}) + \frac{\partial F_{41}}{\partial X_4} = -j(\frac{\partial B_2}{\partial z} - \frac{\partial B_3}{\partial y}) - j\frac{1}{c}\frac{\partial E_1}{\partial (ct)}$$

then

$$\frac{\partial B_3}{\partial y} - \frac{\partial B_2}{\partial z} - \frac{1}{c^2}\frac{\partial E_1}{\partial t} = \mu_0 I_1$$

or

$$\frac{\partial H_3}{\partial y} - \frac{\partial H_2}{\partial z} - \frac{\partial D_1}{\partial t} = I_1$$

Similarly, $v = 2,3$ correspond the other two component equations, hence the second formula of Eq. (10.22) is proved.

If $v = 4$ in Eq. (10.15), we have

$$-\mu_0 c\rho = \frac{\partial F_{\mu 4}}{\partial X_\mu^+} = -j(\frac{\partial F_{14}}{\partial X_1} + \frac{\partial F_{24}}{\partial X_2} + \frac{\partial F_{34}}{\partial X_3}) = -\frac{1}{c}(\frac{\partial E_1}{\partial x} + \frac{\partial E_2}{\partial y} + \frac{\partial E_3}{\partial z}) = -\frac{1}{c}\nabla \cdot \vec{E}$$

which satisfies $\nabla \cdot \vec{D} = \rho$, thus the first formula of Eq. (10.22) is proved.

Then we prove Eq. (10.24) from Eq. (10.23).

If $\mu, \nu, \lambda$ may be equal to 1, or 2, or 3, but not 4 in Eq. (10.23), for example, let $\mu = 2, \nu = 3, \lambda = 1$, then we obtain

$$j(\frac{\partial F_{23}}{\partial X_1} + \frac{\partial F_{31}}{\partial X_2} + \frac{\partial F_{12}}{\partial X_3}) = j(\frac{\partial B_1}{\partial x} + \frac{\partial B_2}{\partial y} + \frac{\partial B_3}{\partial z}) = j\nabla \cdot \vec{B} = 0$$

which is the second formula of Eq. (10.24). If one of the value of $\mu, \nu, \lambda$ is equal to 4, as $\mu = 2, \nu = 4, \lambda = 1$, then we obtain

$$j(\frac{\partial F_{24}}{\partial X_1} + \frac{\partial F_{41}}{\partial X_2}) + \frac{\partial F_{12}}{\partial X_4} = \frac{1}{c}(\frac{\partial E_2}{\partial x} - \frac{\partial E_1}{\partial y}) + \frac{1}{c}\frac{\partial B_3}{\partial t} = 0$$

which is one of components of the first formula of Eq. (10.24), the compound of each component is the first formula.

## 11. Conclusion

We depict Feynman diagram in Minkowski space, it makes the geometric correlation of timelike region and lightlike region express the coupling relation between Dirac particles / antiparticles and electromagnetic field. By using four Dirac spinor equations in Eq. (9.3), we deduce Yang-Mills equation and Maxwell equation, which is the theoretical basis for the geometric express of relativistic quantum mechanics and quantum field theory.

**Figures:**

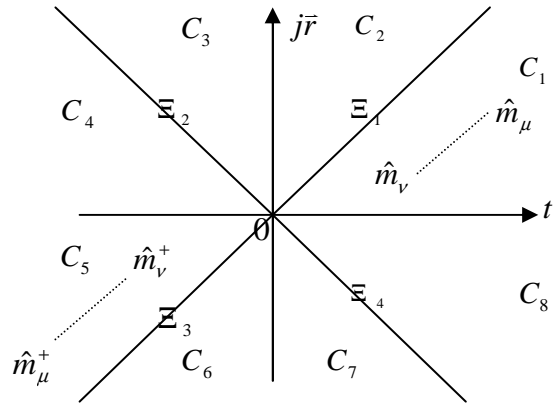

Fig. 1. Geometric point of the particles and antiparticles in Minkowski space

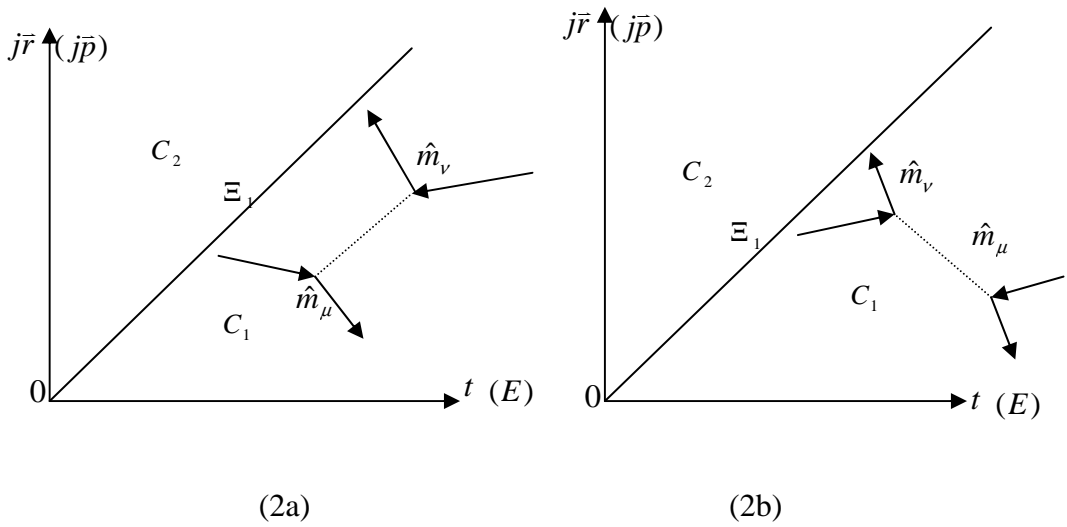

          (2a)                            (2b)

Fig. 2. Feynman diagram in four-dimensional Minkowski space

(a) link line of particles and antiparticles parallel to $\Xi$

(b) link line of particles and antiparticles perpendicular to $\Xi$